%% file: main.tex
\documentclass{article}
\usepackage{spconf,amsmath,graphicx}
\usepackage{subcaption}
\usepackage{booktabs}
\usepackage{xcolor}
\usepackage{graphicx}
\usepackage{lipsum}  
\usepackage{xspace}

\usepackage{enumitem}  

\input{notations}

\title{AudioSlots: A slot-centric generative model for audio separation}
\twoauthors
 {Pradyumna Reddy\sthanks{Work done while interning at Google.}}
	{University College London}
 {Scott Wisdom, Klaus Greff, John R.~Hershey, Thomas Kipf}
	{Google Research}

\begin{document}
\ninept
\maketitle
\begin{abstract}
In a range of recent works, object-centric architectures have been shown to be suitable for unsupervised scene decomposition in the vision domain. Inspired by these methods we present AudioSlots, a slot-centric generative model for blind source separation in the audio domain. AudioSlots is built using permutation-equivariant encoder and decoder networks. The encoder network based on the Transformer architecture learns to map a mixed audio spectrogram to an unordered set of independent source embeddings. The spatial broadcast decoder network learns to generate the source spectrograms from the source embeddings. We train the model in an end-to-end manner using a permutation invariant loss function. Our results on Libri2Mix speech separation constitute a proof of concept that this approach shows promise. We discuss the results and limitations of our approach in detail, and further outline potential ways to overcome the limitations and directions for future work.

\end{abstract}
\begin{keywords}
Speech separation, object-centric representation
\end{keywords}
\input{01_Introduction}

\input{includes/pipeline}
\input{02_RelatedWork}
\input{03_Method}
\input{04_Experiments}
\input{05_Conclusion}

\bibliographystyle{IEEEbib}
\bibliography{main}

\end{document}

%% file: notations.tex
\newcommand{\algoname}{AudioSlots\xspace}

%% file: 01_Introduction.tex
\section{Introduction}


Recently there has been a lot of research into neural network based architectures that operate on set-structured data and architectures that learn to map from unstructured inputs to set-structured output spaces. In particular, in the vision domain, \textit{slot-centric} or \textit{object-centric} architectures underpin recent advances in object detection~\cite{carion2020end} and unsupervised object discovery~\cite{greff2019multi,locatello2020object}.

These \textit{object-centric} architectures have an inbuilt inductive bias of permutation equivariance, making them a natural fit for the task of audio separation.
In this paper we apply the core ideas from these architectures to the problem of sound separation: the task of separating audio sources from mixed audio signals without access to privileged knowledge about the sources or the mixing process. Sound separation is inherently a set-based problem, as the ordering of the sources is arbitrary.

We frame sound separation as a permutation-invariant conditional generative modeling problem: we learn a mapping from a mixed audio spectrogram to an unordered \textit{set} of independent source spectrograms. Our method, AudioSlots, separates audio into individual latent variables per source, which are then decoded into individual source spectrograms. It is built using permutation-equivariant encoder and decoder functions based on the Transformer architecture~\cite{vaswani2017attention} and thus invariant to the ordering of the source latent variables (``slots'').

To evaluate the promise of such an architecture, we train AudioSlots using a matching-based loss to generate separate sources from the mixed audio-signal. We demonstrate our method on a simple two-speaker speech separation task from Libri2Mix~\cite{cosentino2020librimix}. 

While our results primarily constitute a proof of concept for this idea, we find that sound separation with slot-centric generative models shows promise, but comes with certain challenges: the presented version of our model struggles to generate high-frequency details, relies on heuristics for stitching independently predicted audio chunks, and still requires ground-truth reference audio sources for training. We are optimistic that these challenges can be overcome in future work, for which we outline possible directions in this paper.


%% file: includes/pipeline.tex
\begin{figure*}[htp!]
\centering
  \includegraphics[width=0.9\linewidth]{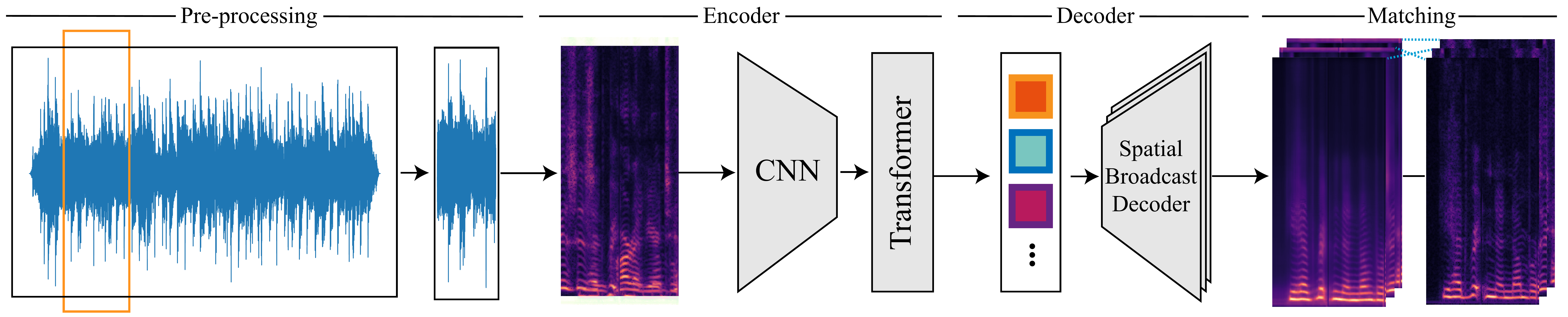}
  \caption{\textbf{Architecture overview.} The input waveform is first cropped and transformed into a spectrogram. Then the neural network encodes the spectrogram to a set of permutation invariant source embeddings $s_{1...n}$, these embeddings are decoded to generate a set of individual source spectrograms. The whole pipeline is supervised with the groundtruth source spectrograms using a matching based permutation invariant loss function. }
\end{figure*}

%% file: 02_RelatedWork.tex
\section{Related Work}
Our work explores a novel set-based generative modeling approach for sound separation. In the following, we provide a brief overview on recent prior learning-based approaches for sound separation as well as on set-based (or \textit{slot-centric}) neural network architectures used in other domains.

\textbf{Sound separation.}\,
A variety of neural network methods have been proposed for supervised sound separation, differing in terms of their sound reconstruction mechanisms and their overall architectures.   
\textit{Mask-based} methods reconstruct sound by predicting separation masks that are applied to an analysis/synthesis basis representation of the input audio signal (such as STFT or learned basis), e.g.~\cite{hershey2016deep, isik2016single,kolbaek2017multitalker,luo2019conv,manilow2019cutting,luo2020dual,subakan2021attention,li2022efficient}.
Alternatively, \textit{direct reconstruction} methods estimate the source signals or their spectra without explicitly estimating masks \cite{tagliasacchi2020seanet,wang2022tf}.

Many generic architectures have been proposed for sound separation, including recurrent networks \cite{weninger2014discriminatively}, convolutional networks \cite{luo2019conv}, U-nets \cite{tagliasacchi2020seanet}, attention networks \cite{koizumi2021df}, and their combinations.  These address the arbitrary permutation of output sources using a permutation-invariant loss during training \cite{isik2016single, kolbaek2017multitalker}.

Some methods have gone further to address permutation invariance at the architecture level by producing un-ordered representations corresponding to each source.  Deep clustering and deep attractor networks \cite{hershey2016deep, chen2017deep, isik2016single}, employ a permutation-equivariant architecture, which operates in an attention-like way over embeddings of each time-frequency bin.     


Our approach produces an embedding for each source, using a slot-based attention mechanism, and decodes it to directly estimate the source spectrogram, using a NeRF-like~\cite{mildenhall2021nerf} method.  This differs from mask-based methods in using direct prediction, and the method of direct prediction using NeRF architecture is novel.  The slot-based attention mechanism is different from previous attention networks for sound separation, and is more closely related to deep clustering. However our attention method works on higher level spectrogram regions rather than individual time-frequency bins, and uses general-purpose attention mechanisms instead of simple affinity-based methods.  



Recent unsupervised approaches, such as mixture invariant training (MixIT) \cite{wisdom2020unsupervised}, us only audio mixtures for training. While we only explore supervised sound separation using ground-truth isolated reference sources in our work, a training setup like in MixIT is orthogonal to our approach and would be interesting to explore in future work.

\textbf{Slot-centric neural networks.}\,
Neural networks that operate on unordered \textit{sets} of features have been studied for some time~\cite{vinyals2015order,zaheer2017deep,lee2019set}. Most closely related to our work are approaches that \textit{generate} an unordered set of outputs conditioned on some input data~\cite{zhang2019deep,kosiorek2020conditional,carion2020end} and methods that use a set of latent variables (``\textit{slots}'') to model permutation-invariant aspects of the input~\cite{greff2019multi,locatello2020object}, such as objects in visual scenes. We refer to the latter as slot-centric neural networks. 
In the context of vision, this class of models forms the basis for many modern scene understanding approaches, including  object-detection~\cite{carion2020end}, panoptic segmentation~\cite{zhou2022slot}, and unsupervised object discovery~\cite{greff2019multi,locatello2020object, reddy2022search}. In our work, we demonstrate that slot-centric generative models similarly hold promise for compositional, permutation-invariant tasks in audio processing, specifically for separating individual audio sources from a mixture.

%% file: 03_Method.tex
\section{Method}
We present a generalized permutation invariant training framework for supervised sound separation. 
Unlike previous methods, we approach the source-separation task from a generative perspective.
The main objective of our method is to project the input audio into a set of embeddings each representing a different source in the input.
These embeddings are then used to generate the magnitude spectrograms of individual sources in a permutation invariant manner.
The whole pipeline is supervised with the ground-truth source spectrograms using a permutation-invariant loss. In rest of this section we elaborate on different steps in our training pipeline. 

\textbf{Preprocessing:}  Given a mixture waveform during training we first randomly crop a 0.5-second audio clip. Then following~\cite{xu2022masked} we transform the clip into a spectrogram using a short-time Fourier transform with window size 512 and hop size 125. Then the absolute values of the complex spectrograms are non-linearly scaled by exponentiating them by power of 0.3 to emphasize low values. These scaled absolute values are passed as the input to the next step.

\textbf{Encoding:} To infer source embeddings, the input spectrogram is first encoded using a ResNet-34 network~\cite{he2016deep} to a $32\times8$ grid of encoded ``image'' features $z$. We use a reduced stride in the ResNet root block to retain a higher spatial resolution. Next a transformer with 4 layers maps $z$ to source embeddings $s_{1...n}$ where $n$ is the number of sources and $s_i \in R^d$.
Unlike the original formulation~\cite{vaswani2017attention}, in each transformer layer we first perform a self-attention operation between query vectors $q$ and then perform cross-attention between the outputs of the self-attention step and $z$. We use same variables as both the key and value in self-attention and cross-attention steps.
The initial queries $q$ of dimensionality $4096$ are learned via backpropagation, similar to DETR~\cite{carion2020end}.

\textbf{Decoding:} 
We use a spatial broadcast decoder~\cite{watters2019spatial} to generate individual source spectrograms from $s_{1...n}$. 
First each embedding $s_i$ are copied across a 2D grid to create tensor $g_i$ of shape $F \times T \times d$ where $F, T$ are the frequency bins and timesteps of the output spectrogram. Then positional embeddings with fourier features \cite{tancik2020fourier} are appended to $g_i$ making its shape $F \times T \times (d+e)$ where $e$ is the size of each fourier feature embeddings.  
Subsequently a fully connected network with shared parameters is applied across all the vectors in $g_i$ to arrive at a set of spectrograms each with shape $F \times T$. Note that this decoder is similar to a NeRF model~\cite{mildenhall2021nerf}, which similarly takes positional codes as input and learns a fully-connected network to produce coordinate-dependent outputs. In our case, we directly produce spectrogram values, arranged on a 2D image grid, as outputs and further condition the network on the respective latent source embeddding $s_i$ for each generated spectrogram.

\textbf{Objective:} The predicted spectrograms are generated in no pre-determined ordering as the source separation problem is permutation invariant. Therefore we need to match estimated spectrograms with ground-truth spectrograms to calculate the loss of the network. 
Among all the possible matches between ground-truth and estimated spectrograms we seek to find the optimal assignment with minimum reconstruction error. We suggest to use the Hungarian matching algorithm to solve this assignment problem because of its speed, accuracy and ability to handle large numbers of items efficiently (although we only use datasets with a small number of sources in our experiments). It works by assigning costs or weights to each possible pair and then selects pairs which minimize the total cost or weight associated with them. 
Finally the mean squared error between matched ground-truth and estimated spectrograms is minimized as the training objective to optimize the network parameters.

\textbf{Source Separation:} \label{sec:seperation_algo}
During testing, given an input mixture waveform we first break it into multiple non-overlapping waveforms of length 0.5-seconds. We add zero-padding at the end in case the input waveform length is not evenly divisible into chunks of 0.5s. These waveforms are preprocessed as mentioned above and are passed as inputs to the network, trained using the above pipeline to estimate absolute values of the spectrograms of individual sources. The estimated spectrograms are first rescaled by exponentiating them by power of 1/0.3 to invert the scaling done during preprocessing. These rescaled estimates are used to calculate the masks in the oracle method to create complex spectrogram estimates of individual sources from the input. Given input spectrogram $I$ the output source spectrograms are computed as $m_i*I$ where $m_i$ is the mask corresponding to the $i^{th}$ source estimated using the spectrograms predicted by the neural network.
These complex spectrograms are inverted to waveforms using an inverse short-time Fourier transform (STFT) and then stitched together, resolving matching using the best match with the ground-truth signal for simplicity.

\textbf{Training:} We train using Adam~\cite{kingma2014adam} for 300k steps with a batch size of 64, a learning rate of 2e-4, 2500 warmup steps and a cosine decay schedule.

%% file: 04_Experiments.tex
\section{Experiments}
\input{includes/reconstruction_comparision}

\input{includes/sisnr_table}

We evaluate the performance of our method on speech separation using the Libri2Mix~\cite{cosentino2020librimix} dataset. We use the anechoic version of the dataset. Each instance in the dataset is sampled at 16kHz and 10 seconds long. Libri2Mix contains contains utterances from both male and female speakers drawn from LibriSpeech~\cite{panayotov2015librispeech}. The train-360-clean split of the dataset contains 364 hours of mixtures and the sources are drawn without replacement.

As mentioned above we use the masking to estimate the complex spectrograms of the individual sources using the input spectrogram and the network predictions. There are various masking functions that can be used ~\cite{7178061}. In our experiments we use the ideal binary mask (IBM)~\cite{li2009optimality} and Wiener filter like mask as mask functions which are defined as: 
\[
\text{IBM:}~m_i=
\begin{cases}
  1 & i=\mathrm{argmax}(m) \\
  0 & otherwise
\end{cases}
\]
\[\text{“Wiener like”:}~m_i= \frac{(m_i)^2}{\Sigma_{i=1}^n (m_i)^2}\]

\textbf{Metrics:} We measure the separation performance using scale-invariant signal-to-noise ratio (SI-SNR)~\cite{le2019sdr} and SI-SNR improvement (SI-SNRi). Let $y$ denote the target and $\hat{y}$ denote the estimate obtained by our method. Then SI-SNR measures the fidelity between $y$ and $\hat{y}$ within an arbitrary scale by rescaling the target: 

\[\text{SI-SNR}(y, \hat{y})=10 \log_{10}\frac{||\alpha y||^2}{||\alpha y - \hat{y}||^2}\]
where $\alpha = \mathrm{argmin}_a ||ay - \hat{y}||^2 = y^T\hat{y}/ ||y||^2$. The SI-SNRi is the difference between the SI-SNR of each source estimate after processing and the SI-SNR obtained using the input mixture as the estimate for each source. During evaluation we first match the targets and estimates to maximize SI-SNR and then average the resulting SI-SNR and SI-SNRi scores. 

\textbf{Results:} In Table~\ref{tbl:comparison} we compare our performance with an autoencoder variant of our method (which receives ground-truth reference sources as input) and the performance of separation obtained using the (preprocessed) ground-truth signals. The metrics computed using the ground-truth signals represent the maximum values that can be obtained with our (lossy) preprocessing. In the autoencoder variant we train the network to reconstruct the individual source signals with $n=1$. We then use the individual estimates as the masks over the complex spectrogram of the mixture. Since the spectrograms contain high-frequency features, this would help us understand the ability of our architecture to faithfully represent these features. We also present an ablation by increasing the crop length in the preprocessing step to 1 second.

The difference in performance between the autoencoder variant and the separation model is only $0.18\pm0.01$. This indicates that our method, AudioSlots, is able learn speech separation well, closely approaching the performance of the baseline which receives fully-separated sources as input.

Still, there is substantial headroom for improvement, both in terms of our model as well as our overall pipeline: prior masking-based approaches~\cite{luo2020dual,subakan2021attention,li2022efficient} already solve Libri2Mix speaker separation to an impressive degree, achieving significantly higher SI-SNR values than we report here.
Very recently, diffusion-based approaches have also shown competitive performance \cite{lutati2023separate}.
This gap is in part due to our lossy preprocessing pipeline: for example, computing STFTs on pre-chunked audio (done here for simplicity) introduces border artifacts which even reduces our ground-truth SI-SNR scores below what other models can achieve. We further zero-pad all audio signals to the same length for simplicity.
Also, since we create masks from the generated spectrograms, we are also bound by the limitations of mask-based methods, e.g.\ that masked spectral content cannot be regenerated.

\textbf{Limitations:} Our experimental comparison highlights the main limitation of our method, which is reconstruction fidelity: both the autoencoder baseline and our AudioSlots model encode the signal using a latent bottleneck representation and tend to discard certain high-frequency details (among others). This is also qualitatively visible in Figure \ref{fig:comb_img}. The comparison further shows that the crop length affects the separation performance: while this can be partially explained by boundary artifacts due to our simplistic cropping approach, it also hints towards sensitivity of AudioSlot's ability to both separate and reconstruct audio spectrograms on chunk length.

\textbf{Discussion:} 
Our results show promise for addressing audio separation using slot-centric generative models that represent the audio using a set of source-specific latent variables. This is a significant departure from earlier methods that directly operate on the input audio using e.g.~a masking-based approach~\cite{wisdom2020unsupervised}. Learning source-specific latent variables further has the benefit that these decomposed latent variables can likely be used not just for generation, but also for recognition tasks, similar to how slots in object-centric computer vision models serve as a basis for object detection and segmentation.

We are optimistic that the current limitations of our approach can be overcome in future work:
\begin{itemize}[leftmargin=*]
    \item To address the issue of reconstruction fidelity (blurry reconstructions for high-frequency features), it is likely that moving away from a deterministic feedforward decoder to e.g.~an autoregressive decoding approach, as in AudioLM~\cite{borsos2022audiolm}, or an iterative diffusion-based decoder, as in~\cite{hawthorne2022multi}, can bridge the gap to high-fidelty generation.
    \item At present, AudioSlots assumes supervision in the form of ground-truth sources. An extension to fully-unsupervised training on raw, mixed audio would be desirable. To this end, we explored replacing the Transformer in AudioSlots with a Slot Attention~\cite{locatello2020object} module which has an inductive bias towards decomposition that allows it to be trained unsupervised in the context of \textit{visual} scene decomposition. In initial experiments we found, however, that this inductive bias might not suffice for decomposing audio spectrograms in a fully-unsupervised fashion. A supervised version of AudioSlots with a Slot Attention module, however, performed similar to the version with a Transformer module in initial experiments, highlighting that further exploration is still promising for future work.
    \item We think that the limitation of processing individual chunks in isolation, which requires post-hoc stitching, can be overcome by using a sequential extension of the model, where slots of the past time step are used as initialization for the next time step as in Slot Attention for Video~\cite{kipf2022conditional}. We leave this for future work.
\end{itemize}

%% file: includes/reconstruction_comparision.tex
\begin{figure*}[t!]
    \centering
  \includegraphics[width=0.92\linewidth]{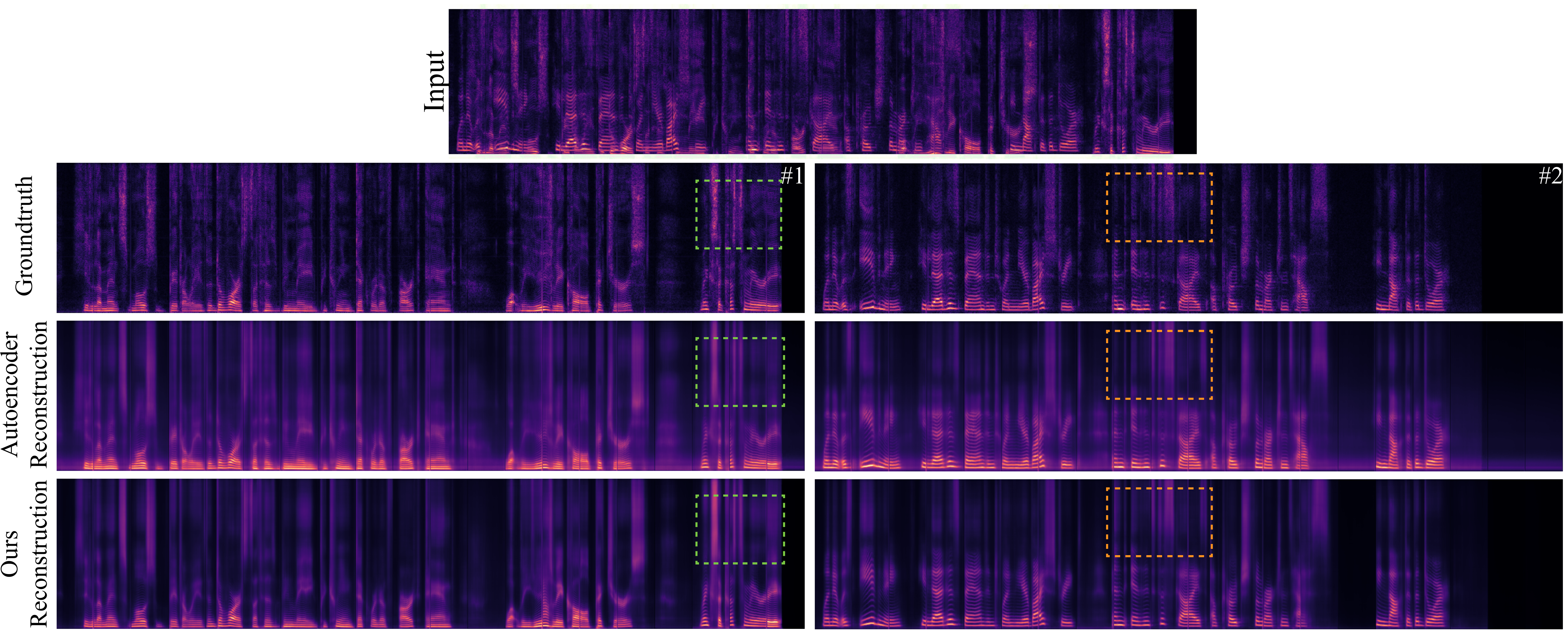}
    \caption{\label{fig:comb_img}
    Comparison between absolute value of the individual source spectrograms of Groundtruth, Autoencoder estimates and AudioSlots (Ours) estimates. The input spectrogram (\textbf{top}) is a mixture and rest of the rows show the spectrograms of the individual sources. The input and groundtruth spectrograms are preprocessed using the steps mentioned in Sec \ref{sec:seperation_algo}. Notice that our method is able to reconstruct harmonics fairly well, however struggles with estimating the high-frequency features (see highlighted example regions).
    }
    \vspace{-10pt}
\end{figure*}

%% file: includes/sisnr_table.tex
\begin{table}[t!]
  \centering
  \caption{\label{tbl:comparison}
    {\bf Results.}  We use the estimated absolute spectrograms as masks over the input complex spectrogram to produce complex spectrograms of the individual sources~\cite{7178061}. In the table below we present SI-SNR[dB] and SI-SNRi[dB] values with IBM and Wiener-like masking. Higher is better. Notice that there is only a minimal drop in performance between Autoencoder and AudioSlots showing that our pipeline is able to learn to separate speech well.}
  \resizebox{\linewidth}{!}{
  \begin{tabular}{r c c c c c}
    \toprule
  & SI-SNR & SI-SNRi  & SI-SNR  & SI-SNRi  \\
  Masking Type& IBM & IBM & Wiener & Wiener \\
  \midrule
    Oracle & 12.07 & 12.07 & 12.32 & 12.33\\  
    Autoencoder & 10.19 & 10.20 & 10.15 & 10.15\\
    AudioSlots & 09.50 & 09.50 & 09.96 & 09.97\\

  \midrule
    Oracle (1-sec) & 13.15 & 13.16 & 13.46 & 13.47\\  
    AudioSlots (1-sec) & 09.66 & 09.67 & 10.20 & 10.20\\
    \bottomrule
  \end{tabular}
  }
\end{table}

%% file: 05_Conclusion.tex
\section{Conclusion}

We present \algoname, a slot-centric generative architecture for audio spectrograms. We demonstrate a proof of concept that \algoname holds promise for addressing the task of audio source separation using structured generative models. While our current implementation of \algoname has several limitations, including low reconstruction fidelity for high-frequency features and requiring separated audio sources as supervision, we are optimistic that these can be overcome and outline several possible directions for future work.